\documentclass[reprint,a4paper,english,showpacs,prd]{revtex4-1}

\usepackage{graphicx}
\usepackage{amssymb}
\usepackage{longtable}
\usepackage{dcolumn}
\usepackage{babel}
\usepackage[latin1]{inputenc}

\usepackage{amsmath,amssymb}

\begin{document}

\title{Black holes and wormholes in AdS branes}

\author{C. Molina}
\email{cmolina@usp.br}
\affiliation{Escola de Artes, Ci\^{e}ncias e Humanidades, Universidade de
  S\~{a}o Paulo\\ Av. Arlindo Bettio 1000, CEP 03828-000, S\~{a}o
  Paulo-SP, Brazil}

\author{J. C. S. Neves}
\email{juliano@fma.if.usp.br}
\affiliation{Instituto de F\'{\i}sica, Universidade de S\~{a}o Paulo \\
C.P. 66318, 05315-970, S\~{a}o Paulo-SP, Brazil}

\begin{abstract}
In this work we have derived a class of geometries which describe black holes and wormholes in Randall-Sundrum-type brane models, focusing mainly on asymptotically anti-de Sitter backgrounds. We show that by continuously deforming the usual four dimensional vacuum background, a specific family of solutions is obtained. Maximal extensions of the solutions are presented, and their causal structures are discussed.
\end{abstract}


\maketitle

\section{Introduction}

The idea of extra dimensions has been attracting much attention in the literature, in part due to the advent of string theory. Strings are promising candidates for a unified description of gravitation and standard model gauge interactions. They require extra dimensions for consistency. Therefore, one problem which has to be addressed by this approach is the observation of only three spacial dimensions. Brane world models provide concrete scenarios where extra dimensions are made compatible with the observed four dimensional universe.

The basic brane model set up is a four dimensional brane, our Universe, immersed in a larger manifold, the bulk. It is generally postulated that the usual matter fields are confined in the brane, while gravitational waves are free to propagate into the bulk. Explicit constructions were proposed by several authors, among them Randall and Sundrum \cite{Randall-Sundrum}. In the Randall-Sundrum (RS) approach, the brane is noncompact and the bulk is anti-de Sitter (AdS). Their specific models were proposed as cosmological scenarios, and little could be said about local objects. Nevertheless, the basic RS models can be expanded so to encompass compact objects, such as black holes (see for example \cite{Maartens} and references therein).

Black holes are important candidates for sources of gravitational radiation. They offer possible observational signatures of Einstein gravity extensions, as discussed, for example, in \cite{Gregory,Seahra,Abdalla,Vitor,Pani,Molina1}. The current and upcoming gravitational wave experiments and the possibility of detecting black holes in accelerators are a strong motivation for the investigation of such models \cite{Cheung,Berti}. Although extensively studied in the context of the usual general relativity, the treatment of such compact objects in RS framework is much more involved. For instance, there are few exact complete solutions defined both in the brane and in the bulk. Stability is also an issue. For example, the black string \cite{black_string}, one of the simplest generalizations of the Schwarzschild geometry, is marked by the so-called Gregory-Laflamme instability \cite{Gregory-Laflamme}. 

Lacking many satisfactory bulk solutions for compact objects in Randall-Sundrum scenarios, one alternative is to search interesting geometries in the brane and invoke Campbell-Magaard-type theorems \cite{Campbel,Seahra2,Casadio2} which guarantee their extensions through the bulk (locally at least). This approach was carried out by several authors, and it will be explored in the present work. In this context, it is particularly convenient to use the formalism developed by Shiromizu, Maeda and Sasaki \cite{Shiromizu}, where gravitational phenomena in a brane immersed in a five dimensional bulk are described by effective four-dimensional Einstein equations. 

Most of the work concerning black holes in extra dimensional scenarios deals with asymptotically flat geometries on the brane \cite{Dadhich_Maartens,Casadio,Bronnikov}. At the present time, not so interesting from the astrophysical point of view, anti-de Sitter solutions have recently gained attention since the AdS/CFT breakthrough. This correspondence ignited the search for a better understanding of negative cosmological spaces and new proposals of gauge/gravity correspondences. Exact AdS solutions play an important role in this program and in this spirit anti-de Sitter branes have been investigated (for example \cite{Diaz,Shiromizu2,Galfard}). We will present a new exact class of spherically symmetric metric vacuum geometries in a Randall-Sundrum brane, emphasizing nonasymptotically flat solutions. And since we are mainly interested in AdS black hole and wormhole geometries, we are led to the solutions with a negative cosmological constant.

The structure of the paper is as follows. In Sec.II we derive an analytic asymptotically anti-de Sitter family of solutions that satisfy a set of specific conditions in the brane. In the approach presented here, we completely fix the arbitrariness in the $g_{tt}$ component of the metric. In Sec.III we investigate the maximal extensions and causal structure of the solutions. Nonextreme and extreme black holes, along with wormholes, are generated. Final remarks are presented in Sec.IV. In this paper we will use the metric signature $diag(-+++)$ and the geometric units $G_{4D} = c =1$, where $G_{4D}$ is the effective four-dimensional gravitational constant.

\section{AdS brane solutions}

One effective four-dimensional formalism for the study of compact objects in a Randall-Sundrum brane world scenario has been developed by Shiromizu, Maeda and Sasaki \cite{Shiromizu}. Within this formalism, the effective four-dimensional vacuum Einstein equations are given by
\begin{equation} 
G_{\mu\nu}=-\Lambda_{4D}g_{\mu\nu}-E_{\mu\nu} \,\, ,
\label{eq_projetada}
\end{equation}
where $G_{\mu\nu}$ is the four-dimensional Einstein tensor associated to the brane metric $g_{\mu\nu}$, $\Lambda_{4D}$ is the brane cosmological constant, and $E_{\mu\nu}$ is proportional to the (traceless) projection on the brane of the five-dimensional Weyl tensor. If we impose staticity and spherical symmetry, we can write the brane metric in the form:
\begin{equation}
ds^{2} = -A(r)dt^{2} + \frac{dr^{2}}{B^{C}(r)}+r^{2} (d\theta^{2} + \sin^{2} \theta d\phi^{2}) \,\, .
\label{metrica_4d}
\end{equation}
The notation $B^{C} = g_{rr}^{-1}$ anticipates results to be derived in this section. A combination of the effective Einstein equations (\ref{eq_projetada}) written without specifying $E_{\mu\nu}$ is the Hamiltonian constraint in the ADM decomposition of the metric, which can be seen as the trace of Eq. (\ref{eq_projetada}):
\begin{equation}
R^{(4)} = 4\Lambda_{4D} = -\frac{12}{L^{2}} \,\, .
\label{Ricci_scalar}
\end{equation}
In Eq. (\ref{Ricci_scalar}), $R^{(4)}$ denotes the four-dimensional Ricci scalar and $L$ denotes the AdS radius. Given a function $A$, we obtain from Eq. (\ref{Ricci_scalar}) a one-parameter family of solutions $\mathcal{B}_{A}$. More specifically, a function $B^{C}\in\mathcal{B}_{A}$ is a solution of the differential equation
\begin{multline}
2 ( 1 - B^{C} ) - r^{2} B^{C} \left\{ \frac{A''}{A} - \frac{(A')^{2}}{2A^{2}} + \frac{A' \left( B^{C} \right)'}{2AB^{C}} \right. \\
+ \left.\frac{2}{r} \left[ \frac{A'}{A}+\frac{\left( B^{C} \right)'}{B^{C}}\right] \right\} = -\frac{12 r^{2}}{L^{2}} \,\, ,
\label{constraint}
\end{multline}
with prime ($'$) denoting differentiation with respect to $r$.

At this level, there is some ambiguity in the choice of $A$. One approach would be to select a specific function $A(r)$ and study the characteristics of the generated family. Natural candidates for $A(r)$ would be the Schwarzschild and Reissner-Nordstr\"om $g_{tt}$ components, which were used in an approach that was carried out in \cite{Casadio}. Another possible approach, developed in \cite{Bronnikov,Bronnikov2}, is to restrict the set of possible alternatives for $A(r)$ based on analyticity considerations at an event horizon. In the present work, we will fix the arbitrariness in the choice of $A(r)$ requiring that the family of solutions generated by (\ref{constraint}) is a continuous deformation of the usual four-dimensional vacuum solutions.

More concretely, Birkhoff-type theorems in usual general relativity uniquely determines the metric functions such that $g_{tt}=-g_{rr}^{-1}$, or in the notation used here, $A=B^{C}$. We propose an extension of the usual case, requiring that for a given choice of $A$, a family of functions $\mathcal{B}_{A}$ obtained must include the function $A$. In other words, if $B^{C}\in\mathcal{B}$, then there is a value for $C$ ($C_{0}$) such that $B^{C_{0}}=A$. We will present in this work the most general solution of Eq. (\ref{constraint}) with this propriety. Therefore we assume the following conditions:
\newtheorem{Condition}{Condition}
\begin{Condition} 
A vacuum in a Randall-Sundrum brane scenario, or equivalently, an effective four-dimensional theory with a traceless stress-energy tensor.
\label{cond1}
\end{Condition}
\begin{Condition} 
Staticity and spherical symmetry.
\label{cond2}
\end{Condition}
\begin{Condition} 
A value for the parameter $C$ ($C=C_{0}$) for which the usual case $A(r)=B^{C_{0}}(r)$ is recovered.
\label{cond3}
\end{Condition}
At this point, the parameter $C$ is a label identifying an element of the family $\mathcal{B}$. Later in this section we will show that $C\in \mathbb{R}$.

With these conditions, the function $A(r)$ is completely determined, and with the specified $A(r)$ it is possible to obtain an analytic expression for the family of functions $\mathcal{B}_{A}$ generated. Indeed, condition \ref{cond1} sets Eq. (\ref{eq_projetada}) as the field equations to the metric, and condition \ref{cond2} leads to Eq. (\ref{constraint}), which relates the metric components $A(r)$ and $B^{C}(r)$. Imposing the extra condition \ref{cond3} on expression (\ref{constraint}), we obtain the following differential equation for $A(r)$:
\begin{equation}
r^{2}A'' +4rA' + 2A = 2 -\frac{12 r^{2}}{L^{2}} \,\, .
\label{eq_diff_A}
\end{equation}
The linear nonhomogeneous differential equation (\ref{eq_diff_A}) has as most general solution, the function
\begin{equation}
A(r) = 1 - \frac{2M}{r} + \frac{Q^{2}}{r^{2}} + \frac{r^{2}}{L^{2}} \,\, ,
\label{A_Q}
\end{equation}
where $M$ and $Q^{2}$ are integration constants in this context.

The zero structure of function $A(r)$ in Eq. (\ref{A_Q}) is relevant for the causal behavior of the family of geometries to be obtained. For the asymptotically anti-de Sitter case this structure is well known. If $L^{2}>0$, $M>0$, and $Q=0$, $A(r)$ has one simple positive real zero ($r_{+}$). With $Q \ne 0$ the result is more complex. In this case, there is an maximum value $Q_{ext}$ for the parameter $Q$ if the function $A$ has positive real zeros. If $L^{2}>0$, $M>0$ and $Q^{2}>0$, the function $A(r)$ has: (i) two simple positive zeros ($r_{+}$ and $r_{-}$) if $|Q|$ is smaller than an extreme value $Q_{ext}$, where $1-Q_{ext}^{2}/r_{+}^{2}+3r_{+}^{2}/L^{2}=0$; (ii) one double positive zero if $|Q|$ is equal to $Q_{ext}$; (iii) no real zeros if $|Q|$ is greater than $Q_{ext}$. If $M\le0$, $A(r)$ does not have any positive zeros, which in this context implies naked singularities. We are not considering this structure, and therefore the positive mass condition $M>0$ will be assumed in the present work.

Fixing the metric component $A(r)$, the next step is to find the associated family of solutions of (\ref{B_Q}) for the function $B^{C}=g_{rr}^{-1}$. We denote this set by $\mathcal{B}_{A}$. The elements $B^{C}(r)\in\mathcal{B}_{A}$ can be analytically determined. They are labeled by the integration constant $C \in \mathbb{R}$ of the first order differential equation (\ref{constraint}). The solution is given by:
\begin{equation}
B^{C}(r) = A(r)\left[1 + \left(C-C_{0}\right) \frac{P(r)}{(r-r_{0})^{k}} \right] \,\, ,
\label{B_Q}
\end{equation}
with
\begin{equation}
P(r) = \frac{L^{4} \exp\left[-\frac{K}{\sqrt{4q-p^{2}}} \arctan\left(\frac{2r + p}{\sqrt{4q - p^{2}}}\right)\right]}{18(r - r_{0-})^{k_{-}}(r^{2} + pr + q)^{2 - (k + k_{-})/2}}\,\,,
\end{equation}
where $r_{0}$ and $r_{0-}$ ($r_{0} > r_{0-}$) are the positive real zeros of the polynomial $f(r)$:
\begin{equation}
f(r) = 3r^{4} + 2L^{2}r^{2} - 3L^{2}Mr + L^{2}Q^{2} \,\, ,
\end{equation}
and the several constants $p,q,k,k_{-},K$ are 
\begin{equation}
p = r_{0}+r_{0-} \,\, , \,\, q = \frac{L^{2}Q^{2}}{3r_{0}r_{0-}} \,\, ,
\end{equation}
\begin{equation}
k = 2 - \frac{r_{0}^{4} + L^{2}Mr_{0} - L^{2}Q^{2}}{r_{0}\left( r_{0} - r_{0-} \right) \left(r_{0}^{2} + pr_{0} + q \right)} \,\, ,
\end{equation}
\begin{equation}
k_{-} = 2 + \frac{r_{0-}^{4} + L^{2}Mr_{0-} - L^{2}Q^{2}}{r_{0-} \left(r_{0} - r_{0-}\right) \left(r_{0-}^{2} +pr_{0-} + q\right)} \,\, ,
\end{equation}
\begin{equation}
K = 2 \left[q\left(\frac{k}{r_{0}} + \frac{k_{-}}{r_{0-}}\right) - p\left(2 - \frac{k + k_{-}}{2}\right)\right] \,\, .
\end{equation}
It is straightforward to check that if $M>0$ and $L^{2}>0$ then: (i) $f(r)$ has two positive real roots if $0<Q<Q_{ext}$, or one positive real root if $Q=0$; (ii) $0\le r_{0-}\le r_{-}\le r_{0}\le r_{+}$; and (iii) $4q-p^{2}>0$. Therefore $P(r)$ is positive definite and the functions $A(r)$ and $B^{C}(r)$ are well defined for $r>r_{0}$. 

The constants $r_{+}$ and $r_{0}$ can be written in terms of $M,Q,L$, although the expressions are somewhat cumbersome. In the limit $Q=0$ the solution assumes a simpler form:
\begin{multline}
\frac{r_{+}}{L} = \left[ \frac{M}{L} + \sqrt{\left(\frac{M}{L}\right)^{2} + \frac{1}{27}}\right]^{1/3} \\
- \frac{1}{3} \left[\frac{M}{L} + \sqrt{\left(\frac{M}{L}\right)^{2} + \frac{1}{27}} \right]^{-1/3} \,\, .
\end{multline}
In this limit, the polynomial $f(r)$ has only one positive real zero $r_{0}$, given by:
\begin{multline}
\frac{r_{0}}{2^{-1/3}L} = \left[\frac{M}{L} + \sqrt{\left(\frac{M}{L}\right)^{2} + \frac{32}{729}}\right]^{1/3} \\
- \left(\frac{32}{729}\right)^{1/3} \left[\frac{M}{L} + \sqrt{\left(\frac{M}{L}\right)^{2} + \frac{32}{729}} \right]^{-1/3} \,\, .
\end{multline}
We also observe that $p\rightarrow r_{0}$, $q\rightarrow L^{2}/3r_{0}$, $r_{0-}\rightarrow0$ and $k_{-}\rightarrow0$ as $Q\rightarrow0$. 

As already commented, the dimensionless parameter $C$ originates as an integration constant of the differential equation (\ref{constraint}). The convention adopted in this work (following \cite{Bronnikov}) is to set $C$ and $C_{0}$ in such a way that if $C=C_{0}$, where
\begin{equation}
C_{0} = \frac{(r_{+} - r_{0})^{k}}{P(r_{+})} \,\, ,
\end{equation}
then $B^{C_{0}}(r)=A(r)$, as required. Therefore, the constraint $C=C_{0}$ gives the Schwarzschild-anti-de Sitter or Reissner-Nordstr\"om-anti-de Sitter geometries, while for $C\ne C_{0}$ we have more complex extensions of the usual four-dimensional vacuum solutions. This remark suggest that $C$ could be interpreted as the parameter which gauges the bulk's influence on the brane. 

Conditions \ref{cond1}-\ref{cond3}, and therefore the solutions given by $A$ and $B^{C}$ in Eqs. (\ref{A_Q})-(\ref{B_Q}), define an effective stress-energy tensor (SET) in the brane $(T^{eff})^{\mu}_{\nu} = - E^{\mu}_{\nu}/8\pi = diag (-\rho^{eff},p_{rad}^{eff},p_{tan}^{eff},p_{tan}^{eff})$. The effective energy density $\rho^{eff}$, effective radial pressure $p_{rad}^{eff}$, and effective tangential pressure $p_{tan}^{eff}$ are given by:
\begin{equation}
- 8\pi \rho^{eff} = -E^{t}_{t} = 
-\frac{3}{L^2} + \frac{B^C - 1}{r^2} + \frac{\left( B^{C} \right)'}{r} \,\, ,
\label{set1}
\end{equation}
\begin{equation}
8 \pi p_{rad}^{eff} = -E^{r}_{r} = 
-\frac{3}{L^2} + \frac{B^C - 1}{r^2} + \frac{A' B^{C}}{r A} \,\, ,
\label{set2}
\end{equation}
\begin{multline}
8 \pi p_{tan}^{eff} = -E^{\theta}_{\theta} = -E^{\phi}_{\phi} \\ 
= +\frac{3}{L^2} - \frac{B^{C}-1}{r^2} - \frac{\left( B^{C} \right)'}{2r} - \frac{A'B^{C}}{2 r A}  \,\, .
\label{set3}
\end{multline}
By construction, the geometries defined by the functions $A(r)=-g_{tt}(r)$ and $B^{C}(r)=g^{-1}_{rr}(r)$ are solutions of the full effective brane equations (\ref{eq_projetada}) with the effective stress-energy tensor given by Eqs. (\ref{set1})-(\ref{set3}). 

An explicit expression for the SET in terms of $M$, $Q$, and $C$ is cumbersome but easily obtained from Eqs. (\ref{set1})-(\ref{set3}). We observe that $\rho^{eff}$, $p_{rad}^{eff}$ and $p_{tan}^{eff}$ are finite for $r>r_{0}$. In general, the weak and null energy conditions are violated, which is consistent with other solutions in brane world scenarios (for example \cite{Casadio,Bronnikov}). In the approach used in this work, we have specified the spacetime metric based on physical and geometrical considerations. But the conditions \ref{cond1}-\ref{cond3} could have been stated in terms of the SET. Paraphrasing these conditions, we may say that the stress-energy tensor given by Eqs. (\ref{set1})-(\ref{set3}) is the unique spherically symmetric and traceless SET which can be obtained by continuously deforming the usual Reissner-Nordstr\"om-anti de Sitter stress-energy tensor.

\section{Maximal extensions}

We now consider Lorentzian, smooth, and simply connected maximal extensions of the obtained solutions. In all extensions constructed here, the AdS character of the solution is seen considering the large $r$ limit. The function $P(r)$ decays to zero as $r\rightarrow\infty$ and the geometry is asymptotically characterized by the pure anti-de Sitter metric:
\begin{equation}
B^{C}(r) = A(r) = 1 + \frac{r^{2}}{L^{2}} + \mathcal{O} \left(\frac{1}{r}\right)\,\, .
\label{asymp_ads}
\end{equation} 
The geometry near spacial infinity therefore have similar structure of the AdS spacetime.

The interior geometries depend of the zero structure of the function $B^{C}(r)$, which in turn depends on the value of the parameter $C$. One important point in the solution (\ref{B_Q}) is that $r_{+}$, the largest zero of $A(r)$, is also the largest zero of the function $B^{C}(r)$ if $C\ge0$ (and the only one, if $C\ge C_{0}$). This implies that the presented metric, in the coordinate system $(t,r,\theta,\phi)$, is well defined in the interval $r \in (r_{+},\infty)$ if $C\ge0$. The maximal extensions in this case are black holes. On the other hand, if $C<0$ there is a zero $r_{thr}$ of $B^{C}(r)$ such that $r_{thr} > r_{+}$, and the maximal extensions are wormhole solutions. 

Since both $A(r)$ and $B^{C}(r)$ have zeros at $r=r_{+}$, we have a candidate for a Killing (and event) horizon. The extensions of the geometries introduced in the previous section will involve interior and exterior blocks (see for example \cite{Walker,Torii}). For completeness, we sketch their construction in the following. Exterior and interior tortoise radial coordinates $r^{\star}$ are introduced
\begin{equation}
\frac{dr^{*}}{dr} = \frac{1}{\sqrt{A(r)B^{C}(r)}} \,\, ,
\label{tartaruga}
\end{equation}
and retarded ($w$) and advanced ($v$) time coordinates are defined in the usual way, $w=t-r^{\star}$ and $v=t+r^{\star}$. From them, coordinates $W$ and $V$ are set as $W=\arctan (w)$ and $V=\arctan (v)$. Finally, timelike and spacelike coordinates $T$ and $X$ are introduced. For example, for the exterior block, $T=(W+X)/2$ and $X=(W-X)/2$. The spacetime metric $g_{\mu\nu}$ (defining $ds^{2}$) is written in terms of a conformally associated metric $\hat{g}_{\mu\nu}$ (defining $d \hat{s}^{2}$) as:
\begin{equation}
d \hat{s}^{2}=\Omega(T,X)^{2}\, ds^{2} \,\, ,
\end{equation}
\begin{equation}
d \hat{s}^{2} = -dT^{2} + dX^{2} + r(T,X)^{2}(d\theta^{2} + \sin^{2} \theta d\phi^{2}) \,\, .
\end{equation}
The physical spacetime, plus singular and infinity limits, are conformally mapped into a manifold with boundary, the associated conformal diagram.

The exterior blocks have the usual AdS form, which can be interpreted as a consequence of the asymptotic limit of $r^{\star}(r)$ tends to a finite value, as seen in Eqs. (\ref{asymp_ads}) and (\ref{tartaruga}). They will be discussed in Secs. \ref{SinBH}, \ref{S-RN-BH}, \ref{NonSinBH}, and \ref{ExtBH}. When a horizon is present, the surface gravity of the black hole $\kappa_{+}$ can be considered. Usually associated to thermodynamic considerations, this quantity is given by:
\begin{equation}
\kappa_{+} = \left\{ 
\begin{array}{lcc}
\sqrt{\frac{C}{2r_{+}^{3}}}\  & \textrm{if} & C>0\\
0                             & \textrm{if} & C=0
\end{array}  \right. \,\, .
\label{surface_gravity}
\end{equation}
If $C<0$, the analytical extension is done for $r<r_{thr}$ and no horizon is present. The spacetime in this case has a wormhole structure. This case will be detailed in Sec. \ref{WH}.

\subsection{AdS singular black holes}
\label{SinBH}

\begin{figure}[tp] 
\begin{center}
\begin{minipage}[c]{0.5\columnwidth}
\includegraphics[width=\columnwidth]{fig1a.eps}
\end{minipage}
\hspace{0.1\columnwidth}
\begin{minipage}[c]{0.3\columnwidth}
\includegraphics[width=\columnwidth]{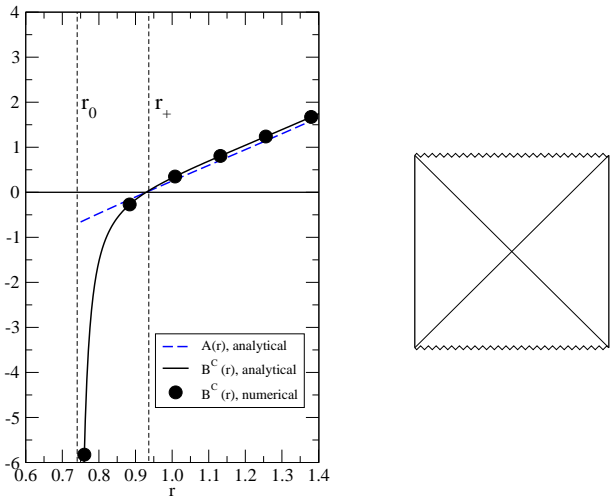}
\end{minipage}
\end{center}
\caption{(Left panel) Typical profiles of the functions $A(r)$ and $B^{C}(r)$ with $C>C_{0}$. In this plot, $M=1$, $Q=1/2$, $L=1$, $C=15$ and $C_0=10.32$. As an illustration, the numerical solution from Eq. (\ref{constraint}) for $B^{C}$ is also shown. (Right panel) Conformal diagram of the maximal extension of the solution with $C>C_{0}$. }
\label{singular_BH}
\end{figure}

If $C>C_{0}$ the only zero of $B^{C}(r)$ in Eq. (\ref{B_Q}) is $r=r_{+}$. At this point it is important to notice that $0<r_{0}<r_{+}$, as we are assuming $M>0$ and $L^{2}>0$. We show typical profiles of the functions $A$ and $B^{C}$ in Fig.\ref{singular_BH} (left panel). It is apparent from Eq. (\ref{B_Q}) that 
\begin{multline}
\lim_{r\rightarrow r_{0}^{+}}B^{C}(r) = -\frac{1}{\left(r-r_{0}\right)^{k}} + \mathcal{O}(r-r_{0})^{0} \rightarrow -\infty \\
\,\, \textrm{with} \,\, r\rightarrow r_{0}\,\, .
\end{multline}
The divergence of the function $B^{C}(r)$ as $r$ tends to $r_{0}$ suggests some peculiarity of the metric in this limit. Indeed, we have a curvature singularity when $r\rightarrow r_{0}$, as seen by the behavior of the Kretschmann invariant:
\begin{multline}
\left| R_{\alpha\beta\gamma\delta}^{(4)}R^{(4)\alpha\beta\gamma\delta}\right| =
\frac{1}{\left( r - r_{0} \right)^{6}} + \mathcal{O}( r - r_{0})^{0}
\rightarrow \infty \,\, \\
\textrm{with} \,\, r \rightarrow r_{0} 
\,\, .
\end{multline}
The effective energy density and pressures are also unbounded. This way, the maximal extension involves the metric functions in the range $(r_{0},\infty)$. Combining the block diagrams of the interior and exterior regions (see \cite{Walker} for example), we obtain the Carter-Penrose diagram of the maximal extension of the solution, with $C>C_{0}$, shown in Fig.\ref{singular_BH} (right panel). We remark that the zero $r_{-}$ does not play any role in the interior causal structure since $r_{-}<r_{0}$, in general.

If $Q=Q_{ext}$ and $C\ge0$, we observe a qualitative change in the causal structure of the spacetime. In this limit, $r_{+}=r_{0-}=r_{0}$, and $r_{+}$ is a (double) zero of the function $A(r)$ as well as a divergence point of the function $B^{C}(r)$. We observe in this case a null singularity and no horizon.

\subsection{Schwarzschild and Reisnner-Nordstr\"om AdS black holes}
\label{S-RN-BH}

With the constraint $C=C_{0}$, we have $B^{C_{0}}(r)=A(r)$. The geometry is equipped with the usual Reisnner-Nordstr\"om anti-de Sitter ($0<Q<Q_{max}$) and Schwarzschild anti-de Sitter ($Q=0$) metrics. In this case, we observe an asymptotically anti-de Sitter background, an event horizon, and an interior Cauchy horizon (if $Q\ne0$). Their Carter-Penrose diagrams are well known (see for example \cite{Torii}), and are not shown in this work.

\subsection{AdS regular black holes}
\label{NonSinBH}

\begin{figure}[tp]
\begin{center}
\begin{minipage}[c]{0.5\columnwidth}%
\includegraphics[width=\columnwidth]{fig2a.eps}
\end{minipage}
\hspace{0.1\columnwidth}
\begin{minipage}[c]{0.3\columnwidth}%
\includegraphics[width=\columnwidth]{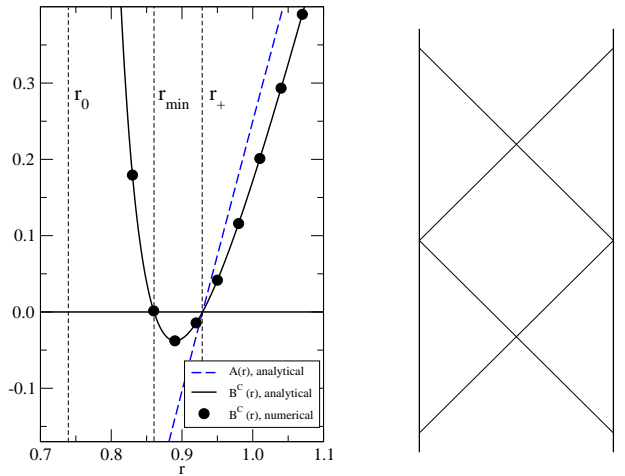}
\end{minipage}
\end{center}
\caption{(Left panel) Typical profiles of the functions $A(r)$ and $B^{C}(r)$ with $0<C<C_{0}$. In this plot, $M=1$, $Q=1/2$, $L=1$, $C=5$ and $C_0=10.32$. As an illustration, the numerical solution from Eq. (\ref{constraint}) for $B^{C}$ is also shown. (Right panel) Conformal diagram of the maximal extension of the solution with $0<C<C_{0}$.}
\label{nonsingular_BH}
\end{figure}

If $0<C<C_{0}$, the causal structure of the interior region changes. For this range variation of the parameter $C$, the function $B^{C}(r)$ has a simple zero at a point $r_{min}$ inside the interval $(r_{0},r_{+})$. We show typical profiles of the functions $A$ and $B^{C}$ in Fig.\ref{nonsingular_BH} (left panel). The analytic extension of the spacetime to a larger Lorentzian manifold implies that the radial coordinate $r$ is valid in the interval $r>r_{min}$. The inside block has a wormhole structure, as seen by the coordinate transformation $(t,r) \rightarrow (t,x)$, where $x=r_{min} + x^{2}$. The black hole interior is regular, with no singularity present. Effective energy density and pressures are bounded. This structure is similar to the one present in the scenarios discussed in \cite{Casadio,Bronnikov}. The Carter-Penrose diagram in this case is shown in Fig.\ref{nonsingular_BH} (right panel).

\subsection{AdS extreme black holes}
\label{ExtBH}

\begin{figure}[tp]
\begin{center}
\begin{minipage}[c]{0.5\columnwidth}%
\includegraphics[width=\columnwidth]{fig3a.eps}
\end{minipage}
\hspace{0.1\columnwidth}
\begin{minipage}[c]{0.2\columnwidth}%
\includegraphics[width=\columnwidth]{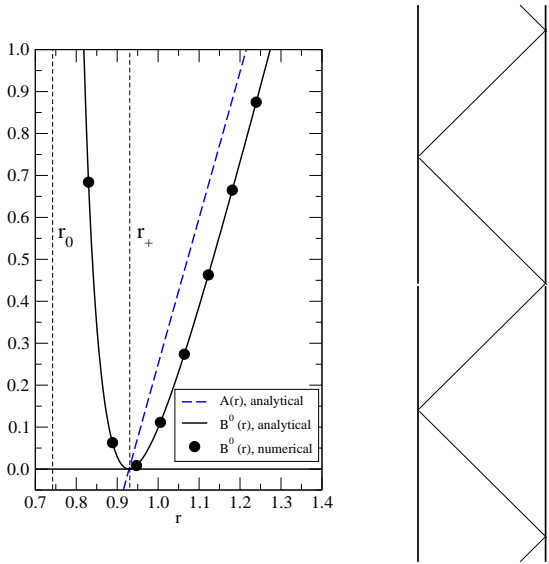}
\end{minipage}
\end{center}
\caption{(Left panel) Typical profiles of the functions $A(r)$ and $B^{0}(r)$ with $C=0$. In this plot, $M=1$, $Q=1/2$, $L=1$, $C=0$ and $C_0=10.32$. As an illustration, the numerical solution from Eq. (\ref{constraint}) for $B^{C}$ is also shown. (Right panel) Conformal diagram of the maximal extension of the solution with $C=0$. }
\label{extreme_BH}
\end{figure}

If $C=0$, the function $B^{0}(r)$ has a double zero at $r=r_{+}$ and it is negative nowhere. We show typical profiles of the functions $A$ and $B^{0}$ in Fig.\ref{extreme_BH} (left panel). The horizon is a double horizon and the surface gravity is zero, as indicated in Eq. (\ref{surface_gravity}).

Indeed, one way to extend the geometry beyond the double horizon is to rewrite the metric in terms of the quasiglobal coordinate $u$, defined by $g_{tt}g_{uu}=-1$,
\begin{equation}
ds^{2} = -\mathcal{A}(u)dt^{2} + \frac{du^{2}}{\mathcal{A}(u)} + r(u)^{2}(d\theta^{2} + \sin^{2}\theta d\phi^{2}) \,\, .
\label{metrica_qn}
\end{equation}
The quasiglobal radial coordinate is convenient in the analysis of this and other spherically symmetric geometries \cite{Bronnikov3}.  In a sense, this coordinate extends to general spherically symmetric geometries the good properties of the usual $r$ coordinate in the Schwarzschild solution context. We will show that the near horizon geometry, described with the coordinate systems based on $u$, is well defined on both sides of the Killing horizon, hence the name ``quasiglobal'' coordinate for $u$ \cite{Bronnikov,Bronnikov3}. 

To check that the spacetime defined by $C=0$ can be analytic extended, we observe that the function $\mathcal{A}(u)$ has a double zero at $u=0$, where $r(0)=r_{+}$: 
\begin{equation}
\mathcal{A}(u) = \frac{1}{2} \frac{\left(2-r_{+}R\right)^{2}}{A'(r_{+})r_{+}^{4}}u^{2}+\mathcal{O}\left(u^{4}\right).
\label{Au}
\end{equation}
The interior extension is possible since the extended function $r(u)$ is analytic. In particular, near the Killing horizon $u=0$:
\begin{equation}
r(u) = r_{+} + \frac{1}{2} \frac{\left(2-r_{+}R\right)^{2}}{A'(r_{+})^{2}r_{+}^{4}} u^{2} + \mathcal{O}\left(u^{4}\right).
\end{equation}

In the coordinate system $(t,u,\theta,\phi)$ it is apparent that the extension has the correct signature in the interior of the black hole ($u<0$), as seen in Eqs. (\ref{metrica_qn}) and (\ref{Au}). Advanced and retarded coordinates can be written in terms of the radial coordinate $u$, following the usual procedure. It is observed that the maximal Lorentzian extension is formed by asymptotically AdS regions, in the limits $u \rightarrow \infty$ and $u \rightarrow -\infty$, separated by a Killing horizon ($u = 0$, $r = r_{+}$). The Carter-Penrose diagram is presented in Fig.\ref{extreme_BH} (right panel).

\subsection{AdS wormholes}
\label{WH}

\begin{figure}[tp]
\begin{center}
\begin{minipage}[c]{0.5\columnwidth}%
\includegraphics[width=\columnwidth]{fig4a.eps}
\end{minipage}
\hspace{0.1\columnwidth}
\begin{minipage}[c]{0.35\columnwidth}%
\includegraphics[width=\columnwidth]{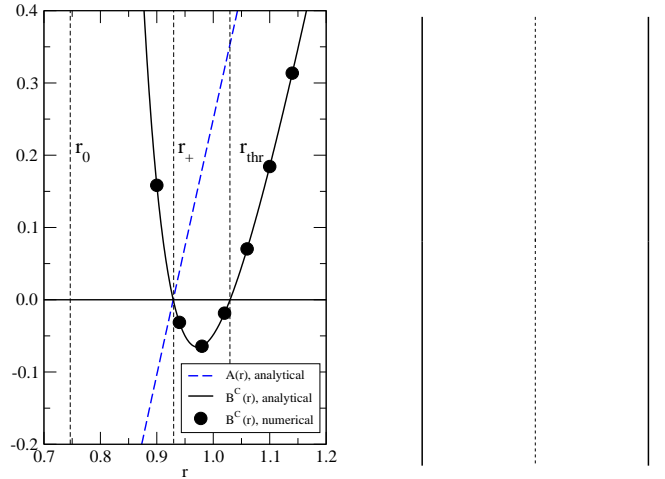}
\end{minipage}
\end{center}
\caption{(Left panel) Typical profiles of the functions $A(r)$ and $B^{C}(r)$ with $C<0$. In this plot, $M=1$, $Q=1/2$, $L=1$, $C=-5$ and $C_0=10.32$. As an illustration, the numerical solution from Eq. (\ref{constraint}) for $B^{C}$ is also shown. (Right panel) Conformal diagram of the maximal extension of the solution with $C<0$. The vertical dashed line represents the wormhole throat.}
\label{wormhole}
\end{figure}

If $C<0$, the function $B^{C}(r)$ has two simple positive zeros, $r_{+}$ and $r_{thr}$, where $r_{+}<r_{thr}$ and $A(r_{thr})\ne0$. We show typical profiles of the functions $A$ and $B^{C}$ in Fig.\ref{wormhole} (left panel). The coordinate system $(t,r,\theta,\phi)$ is not admissible if $r<r_{thr}$.  A possible choice of radial coordinate to cover both regions is the proper length $\ell$ \cite{Bronnikov2,Morris_Torne}, defined as
\begin{equation}
\frac{d\ell}{dr} = \frac{1}{\sqrt{B(r)}} \,\, .
\end{equation}
The function $r(\ell)$ has a minimum $r_{thr}$ at $\ell=0$ (value of $\ell$ is conventionally chosen, without loss of generality). The zero $r_{+}$ does not play any role in the interior causal structure since $r_{+}<r_{thr}$. 

The curvature invariants, effective energy density and effective pressures are everywhere finite in the wormhole. It is straightforward to check that the surface $\ell=0$ in the extended geometry is timelike and transversable. The solution therefore splits into two asymptotically anti-de Sitter regions. The maximal extension is a wormhole connecting two anti-de Sitter branches, with an wormhole throat at $\ell=0$. Its associated conformal diagram is presented in Fig.\ref{wormhole} (right panel).

\section{Final remarks}

We have obtained a family of exact solutions of the effective Einstein equations in a four-dimensional brane. Conditions \ref{cond1}-\ref{cond3}, and therefore the solutions given by Eqs. (\ref{A_Q})-(\ref{B_Q}), define an effective stress-energy tensor $T^{eff}$ in the brane given by Eqs. (\ref{set1})-(\ref{set3}). Fixing the cosmological constant $\Lambda_{4D}$, the solutions are parametrized by three real constants, namely $M$, $Q$ and $C$. Varying these parameters, the maximal extension manifolds describe a large class of objects, including naked singularities, black holes protecting spacelike singularities, regular and extreme black holes, and wormholes. 

In this paper we have focused on asymptotically anti-de Sitter solutions. But relaxing the condition $\Lambda_{4D}<0$ we have asymptotically flat and de Sitter metrics. The solution given by the expressions (\ref{A_Q}) and (\ref{B_Q}) can be adapted for asymptotic flat or de Sitter geometries, but it has distinct characteristics depending on the sign of the four-dimensional cosmological constant $\Lambda_{4D}$. If $\Lambda_{4D}>0$ the geometry is asymptotically de Sitter, and in general it does not describe black holes, since for most values of $C$ we have $r_{0}>r_{+}$ and there is a curvature singularity after the ``would be'' event horizon. If $\Lambda_{4D}=0$ the brane solution is asymptotically flat, and expressions derived here generalize cases already discussed, for example, in \cite{Dadhich_Maartens,Casadio,Bronnikov,Chamblin_Reall}. 

Although we presented the exact solutions within a world model context, they can also be seen as AdS solutions in the usual general relativity. Within this interpretation, the solution presented by Eqs. (\ref{A_Q}) and (\ref{B_Q}) are viewed as the metric generated by the matter content described by a traceless stress-energy tensor $T^{eff}$. Therefore, they are natural candidates on the gravity side of the anti-de Sitter/conformal field theory correspondence.

\begin{acknowledgments}
This work was partially supported by Conselho Nacional de Desenvolvimento Cient\'{\i}fico e Tecnol\'ogico (CNPq) and Coordena\c{c}\~ao de Aperfei\c{c}oamento de Pessoal de N\'{\i}vel Superior (CAPES), Brazil.
\end{acknowledgments}

\end{document}